\documentclass[12pt]{iopart}
\usepackage{xcolor}
\usepackage{graphicx}
\usepackage{cite}
\usepackage{hyperref}
\newcommand{\Hop}{\hat{H}}

\usepackage{lineno}

\begin{document}

\title[Few particles with an impurity in a one-dimensional harmonic trap]{Few particles with an impurity in a one-dimensional harmonic trap}

\author{A. Rojo-Franc\`as\textsuperscript{1,2},
F. Isaule\textsuperscript{3,4}, and
B. Juli\'{a}-D\'{i}az\textsuperscript{1,2}}

\address{$^1$ Departament de F\'isica Qu\`antica i Astrof\'isica,
Facultat de F\'{\i}sica,  Universitat de Barcelona,
Mart\'i i Franqu\`es 1, E–08028 Barcelona, Spain.}
\address{$^2$ Institut de Ci\`encies del Cosmos (ICCUB), Universitat de Barcelona,
Mart\'i i Franqu\`es 1, E–08028 Barcelona, Spain.}
\address{$^3$ School of Physics and Astronomy, University of Glasgow, Glasgow G12 8QQ, United Kingdom.}
\address{$^4$ Instituto de Física, Pontificia
Universidad Católica de Chile,
Avenida Vicuña Mackenna 4860,
Santiago, Chile.}
\ead{bruno@fqa.ub.edu}
\vspace{10pt}
\begin{indented}
\item[]March 2024
\end{indented}

\begin{abstract}
We present a comprehensive study of the static properties of a mobile impurity interacting with a bath with a few particles trapped in a one-dimensional harmonic
trap. We consider baths with either identical bosons or distinguishable particles and we focus on the limiting case where the bath is non-interacting. We provide numerical results for the energy spectra and density profiles by means of the exact diagonalization of the
Hamiltonian, and find that these systems show non-trivial solutions, even in the limit of infinite repulsion. A detailed physical interpretation is provided for the lowest energy states. 
In particular, we find a seemingly universal transition from the impurity being
localized in the center of the trap to being expelled outside the majority 
cloud. We also develop an analytical ansatz and a mean-field solution to compare them with our numerical results in limiting configurations.
\end{abstract}

%
%
\submitto{\NJP}
%
%
%

\section{Introduction}
\label{sec:intro}

The study of impurities interacting with a quantum bath is of relevance in many branches of physics. Examples range from the famous problems of electrons coupled to ionic crystals\cite{landau_effective_1948} and $^3$He impurities in liquid Helium~\cite{girardeau_motion_1961,fabrocini_3mathrmhe_1998} to nucleon impurities in neutron matter~\cite{kutschera_proton_1993}. More recently, ultracold atom experiments~\cite{bloch_many-body_2008,bloch_quantum_2012} have opened a new avenue for probing impurities in quantum baths, offering a unique setting to control impurity-bath systems. In particular, both Fermi~\cite{schirotzek_observation_2009,nascimbene_collective_2009} and Bose~\cite{jorgensen_observation_2016,hu_bose_2016,pena_ardila_analyzing_2019,yan_bose_2020} polarons have already been observed experimentally, and the realization of more sophisticated impurity systems is expected in the near future. 

The mentioned developments have motivated extensive theoretical studies of impurities in a variety of configurations, including different statistics~\cite{vlietinck_quasiparticle_2013,christensen_quasiparticle_2015} and dimensions~\cite{volosniev_analytical_2017,koschorreck_attractive_2012,ardila_strong_2020,hryhorchak_mean-field_2020}. These studies can further test the validity of our theoretical approaches and bridge the gap between few- and many-body physics. Indeed, the important impact of impurity-bath correlations means that perturbative approaches are often not reliable, especially in strongly-interacting regimes, requiring us to rely on more sophisticated theoretical techniques.

In this direction, one-dimensional gases~\cite{giamarchi_quantum_2004,mistakidis_few-body_2023} offer a unique platform to study impurities~\cite{chen_intra-_2022,petkovic_dynamics_2016,grusdt_bose_2017,will_polaron_2021,brauneis_impurities_2021,keiler_doping_2020,massel_dynamics_2013,burovski_mobile_2021,yordanov_mobile_2023}. Fluctuations are enhanced in one-dimension, increasing the importance of impurity-bath correlations. Furthermore, one-dimensional systems often offer us reliable solutions, such as from the Bethe ansatz~\cite{bethe_zur_1931} or from numerical diagonalization of few-particles systems~\cite{raventos_cold_2017,weise_exact_2008}.
In particular, the theoretical study of dressed impurities in bosonic mediums, that is Bose polarons, has received increased attention in one dimension. Bosonic baths often offer richer impurity physics than their fermionic counterparts, which has
motivated recent state-of-the-art studies of one-dimensional Bose polarons in the regime of strong interactions~\cite{grusdt_bose_2017,will_polaron_2021}. However, a comprehensive description has not been yet achieved, with current questions including the dynamic formation and stability of polarons~\cite{will_dynamics_2023}
and the induced correlations between polarons~\cite{dehkharghani_coalescence_2018}.
Experimentally, ultracold atoms have provided realizations of one-dimensional gases for two decades~\cite{paredes_tonksgirardeau_2004,kinoshita_observation_2004}, including the observation of impurity dynamics in a bosonic bath~\cite{catani_quantum_2012}.

Significant theoretical efforts have been put to understand the behavior of impurities in one-dimensional harmonic traps~\cite{dehkharghani_quantum_2015, zinner_fractional_2014,mistakidis_quench_2019,mistakidis_repulsive_2019,wlodzynski_several_2022,mistakidis_many-body_2020,rammelmuller_magnetic_2023,mistakidis_induced_2020,theel_counterflow_2022}
, as they better simulate experimental conditions. Here, we study impurities in one-dimensional harmonic traps by means of the exact diagonalization (ED) technique~\cite{ garcia-march_entanglement_2016} with a renormalization of the interaction to improve the convergence~\cite{rojo-francas_direct_2022}.
Even though ED calculations are restricted to a small number of particles, they provide us with highly accurate results for many properties, including the low-energy spectrum, density profiles and even dynamics, which often cannot be easily obtained with other techniques. Furthermore, ED studies can provide helpful insight into systems with more particles.

In this work, we study a one-dimensional harmonic trap loaded 
with one mobile impurity interacting repulsively with a bath of 
a few (two to seven) non-interacting particles by performing ED calculations~\cite{deuretzbacher_evolution_2007,koscik_optimized_2018,rojo-francas_static_2020}. We consider that the bath's particles are either bosons or distinguishable particles.
We note that a similar configuration as the one considered here 
has been studied previously in~\cite{zinner_fractional_2014}, where one 
particle interacts with a small bath of non-interacting identical 
bosons. In particular, Ref.~\cite{zinner_fractional_2014} provides an analytical result for three particles in the 
limit of infinite repulsion, which is used to benchmark our calculations.
We examine both the regimes of weak and strong interactions.
We compare our numerical calculations with mean-field (MF) solutions in the regime of weak repulsion, while we also propose an ansatz for the ground state in the limit of infinite repulsion. Our ED calculations correctly connect both limits. In particular, in the regime of strong repulsion, we find that the physical properties saturate to non-trivial solutions. These solutions are correctly captured by our analytical ansatz. We also examine the dependence of the energy and density profiles on the number of particles to provide insight into many-body scenarios. 

As shown by similar works~\cite{dehkharghani_quantum_2015}, while for weak repulsion the bath and impurity are localized at the center of the trap, for strong repulsion above a critical interaction strength the bath expels the impurity to the borders of the trap. This impurity-bath separation can be qualitatively understood as if the impurity was trapped in a double well, i.e. the bath particles localized in the center of the trap play the role of the central barrier. We characterize the transition between the two regimes by studying the position of the maximum of impurity density, which goes from zero to a finite value. In addition, we find that the critical interaction strength for this separation shows a universal behavior in baths with a different number of particles.

This work is organized as follows. In \sref{sec:model} we present our model and numerical considerations. We also present the complementary MF approach and the ansatz for infinite repulsion.
In \sref{sec:energy} we examine the energies of the system for a wide range of interaction strengths, with a detailed analysis of both the ground state and the low-energy spectrum.  Afterward, in \sref{sec:density} we examine the ground-state density profiles for a representative set of the number of particles and interaction strengths. Finally, in \sref{sec:concl} we present the main conclusions of our work and an outline for future directions.

\section{Model}
\label{sec:model}

We consider a one-dimensional system with $N$ particles of equal mass $m$ trapped in a harmonic
potential and that interact through short-range potentials. 
We assume that one particle, the impurity, interacts with equal inter-atomic potentials with the other $N_b=N-1$ particles, while the $N_b$ particles in the \emph{bath} do not interact among themselves. As mentioned, we consider baths with either identical bosons or distinguishable particles
By approximating the impurity-bath interaction by an effective contact potential of strength $g$, the Hamiltonian takes the form
\begin{equation}
    \Hop=\sum_{i=1}^N\left[-\frac{\hbar^2}{2m}\frac{\partial^2}{\partial x_i^2}+\frac{m\omega^2}{2}x_i^2\right]+
    g\sum_{i\ne I}^{N_b}\delta(x_I-x_i)\,,
     \label{sec:model;eq:HO}
\end{equation}
where $\omega$ is the harmonic oscillator (HO) trapping frequency and the interaction strength is related to the one-dimensional $s-$wave scattering length $a_{\mathrm{1D}}$ via $g=-2\hbar^2/ma_{\mathrm{1D}} $ (for details see Refs.~\cite{olshanii_atomic_1998,haller_realization_2009,busch_two_1998}). In the rest of this text, we employ lowercase letters to denote arbitrary particles, whereas we employ capital letters to denote a specific particle. From now on, we 
consider repulsive interactions $g>0$, leaving the study of the attractive branch for future work. We note that in one dimension many of the interesting properties of impurities appear in the repulsive branch~\cite{will_polaron_2021}.

We first stress that the solution in the non-interacting limit $g=0$ is simply given by the textbook solution of the harmonic oscillator. In contrast, interacting systems $g\neq 0$ require more careful treatment, such as from a perturbative calculation or a numerical diagonalization. In this work, in a first instance, we perform exact diagonalizations (ED) of the Hamiltonian~(\ref{sec:model;eq:HO}), where we employ many-body basis in the space of single-particle HO modes, which we truncate up to a chosen HO-mode cutoff. Within this truncated subspace, we diagonalize the Hamiltonian using standard numerical routines. We refer to Refs.~\cite{rojo-francas_static_2020,plodzien_numerically_2018} for detailed reviews on the ED method for harmonic traps.

Due to the use of a truncated basis, and despite the name, the results from ED are not exact, and usually need some correction or extrapolation~\cite{ernst_simulating_2011,grining_many_2015}. Naturally, the quality of the results improves when we increase the number of HO modes. However, the number of possible basis states grows rapidly with the number of particles, restricting our calculations to a few particles (less than ten in this work).
Nevertheless, to improve the quality of our results and to access systems with more particles, we renormalize the interaction strengths with the known two-body solution following the approach described in Ref.~\cite{rojo-francas_direct_2022}. With this method, one corrects the value of the interaction strength used in the initial numerics to a more accurate physical one. This method is well
tested in the symmetric cases of a few particles and, as we show later, it provides an excellent agreement with known analytical solutions for one impurity interacting with two bath particles.

\subsection{Mean-field solution}
\label{sec:model;subsec:mf}

To complement the ED calculations, we also study the weakly-interacting regime within a mean-field (MF) approximation. This enables us to gain further insight and examine systems with a large number of particles, which we cannot access with ED. We introduce the MF wavefunction
\begin{equation}
\label{sec:model;eq:wf_s.c.}
    \Psi(x,x_1,...,x_N)=\phi(x)\prod_{i=1}^{N_b}\varphi^i(x_i)\,,
\end{equation}
where $\phi(x)$ is the impurity wavefunction and $\varphi^i(x_i)$ is the $i-th$ bath particle wavefunction. By employing~(\ref{sec:model;eq:wf_s.c.}), we obtain the MF equations
\begin{equation}
\label{sec:model;eq:s.c.imp}
    \left( -\frac{\hbar^2}{2m}\frac{d^2}{d x^2}+\frac{m\omega^2}{2}x^2 +g\sum_i |\varphi^i(x)|^2 \right)\phi(x)= \epsilon_I \phi(x)\,,
\end{equation}
\begin{equation}
\label{sec:model;eq:s.c.bath}
    \left( -\frac{\hbar^2}{2m}\frac{d^2}{d x^2}+\frac{m\omega^2}{2}x^2 +g|\phi(x)|^2 \right)\varphi^i(x)= \epsilon_b \varphi^i(x)\quad i=1,..,N_b\,,
\end{equation}
where $\epsilon_I$ and $\epsilon_b$ are the eigenvalues of both equations and are used as the convergence parameter. We diagonalize Eqs.~(\ref{sec:model;eq:s.c.imp}) and (\ref{sec:model;eq:s.c.bath}) self-consistently until they meet a convergence criteria for $\epsilon_I$ and $\epsilon_b$. We can choose ground- or excited-state solutions depending on the chosen wavefunctions states during the self-consistent calculation.
Naturally, in the ground state, all the wavefunctions of the bath are equal [$\varphi^i(x)=\varphi(x)$], and thus, the interaction term in~\eref{sec:model;eq:s.c.imp} takes the form $g N_b|\varphi(x)|^2$.
In contrast, one excited solution corresponds to having one bath's particle in its first excited state $\varphi_1$, while the rest of the $N_b-1$ particles in the bath are in their ground state $\varphi_0$. In this case, the interaction term in \eref{sec:model;eq:s.c.imp} takes the form $g(N_b-1)|\varphi_0(x)|^2+g|\varphi_1(x)|^2$. Higher excited states behave analogously. 

From the converged solutions for $\phi(x)$ and $\varphi^i(x_i)$, we extract the densities of each particle from $\rho_I(x)=|\phi(x)|^2$ and $\rho_i(x)=|\varphi^i(x)|^2$. Finally, we compute the energy from the functional
\begin{eqnarray}
\label{sec:model;eq:s.c.energy}
     E_{\mathrm{MF}}=&\int \phi^*(x)\left(-\frac{\hbar^2}{2m}\frac{\mathrm{d}^2}{\mathrm{d}x^2}+\frac{m\omega^2}{2}x^2  \right)\phi(x) dx \nonumber \\
     &+\sum_{i= 1}^{N_b}\int \varphi^{i*}(x)\left( -\frac{\hbar^2}{2m}\frac{\mathrm{d} ^2}{\mathrm{d} x^2}+\frac{m\omega^2}{2}x^2 \right)\varphi^i(x) dx \nonumber\\
     &+g\sum_{i= 1}^{N_b}\int|\varphi^i(x)|^2|\phi(x)|^2 dx\,.
\end{eqnarray}
where we perform the derivatives and integrals numerically. We examine the energies in \sref{sec:energy} and the densities profiles in \sref{sec:density}.

\subsection{Ansatz for infinite repulsion}
Finally, we also complement the ED calculations for strong repulsion by proposing an ansatz for the wavefunction in the limit of infinite repulsion. By taking into account that in the limit $g\to\infty$ the impurity cannot be at the same position as any of the bath particles, we introduce an ansatz with Jastrow-like correlations \cite{jastrow_many-body_1955}
\begin{equation}
\label{sec:model;eq:ansatz2}
    \fl \Psi_{N}(x_I,x_i)=\alpha_N \rme^{-x_I^2 / 2\sigma_I^2}\sum_{N_a=0}^{\lfloor N_b/2 \rfloor}\sum_P \prod_{i\ne j}^{2N_a} |x_I-x_i| |x_I-x_j| \rme^{-(x_i^2+x_j^2)/2\sigma_b^2} \prod_{k\ne i,j}^{N_b-2 N_a } (x_I-x_k) \rme^{-x_k^2 / 2\sigma_b^2}\,,
\end{equation}
where $x_I$ is the position of the impurity, $x_b$ are the positions
of the $N_b$ bath particles, $\alpha_N$ is a normalization factor that
depends on the number of particles $N$, and $\sigma_I$ and $\sigma_b$ are variational parameters that can be adjusted numerically. In the sums, $N_a$ denotes the number of 
pairs of absolute values and $P$ are the permutations of the index $i,j$ with the index $k$.

We first note that (\ref{sec:model;eq:ansatz2}) improves over a similarly inspired ansatz proposed in Ref.~\cite{garcia-march_entanglement_2016}. However, the previous ansatz does not contain the terms with absolute values nor the $\sigma$ parameters. The introduction of a linear combination of terms with $(x_I-x_i)(x_I-x_j)$ and  $|x_I-x_i||x_I-x_j|$, as well as the introduction of the parameter $\sigma$, enable us to provide a more accurate variational solution. To understand this, we note that a simple ansatz without the linear combination results in the same energy as one with a different number of pairs of absolute values $N_a$. Indeed, for two particles both
\begin{equation}\label{sec:model;eq:ansatz_N3_1}
    \psi(x_I,x_i)=\alpha (x_I-x_A)(x_I-x_B)\rme^{-(x_I^2+x_A^2+x_B^2)/2}\,,
\end{equation} 
and
\begin{equation}\label{sec:model;eq:ansatz_N3_2}
\psi(x_I,x_i)=\alpha |x_I-x_A| |x_I-x_B|\rme^{-(x_I^2+x_A^2+x_B^2)/2}\,,
\end{equation}
result in the same energy. However, these wavefunctions are not orthogonal; thus, a linear combination of them gives a better variational solution. We have found that the minimum energy is obtained when all the functions have the same weight (see \ref{app:analytical} for more details).

As an example, the ansatz for three particles ($N_b=2$) reads
\begin{equation}\label{sec:energy;eq:ansatz_3p}
   \fl  \Psi_{3}(x_I,x_A,x_B)=\alpha_{3}\left((x_I-x_A)(x_I-x_B)+|x_I-x_A||x_I-x_B|\right) \rme^{-x_I^2/2\sigma_I^2 -\left(x_A^2+x_B^2\right)/2\sigma_b^2}\,,
\end{equation}
where $A$ and $B$ denote the two particles in the bath. In this example, the first term has $N_a=0$, i.e. without absolute values, while the second term has $N_a=1$, i.e. 
a pair of absolute values.
For an increasing number of particles, the ansatz takes analogous, but more complicated forms. We provide examples of the ansatz for four and five particles in~\ref{app:analytical}.

Our goal is to minimize the energy obtained with this ansatz. Therefore, we find the optimal values of $\sigma_I$ and $\sigma_b$ for the chosen number of particles. We report the energies and the $\sigma$ parameters obtained up to $N=8$ in~\ref{app:analytical}. We find that our ansatz provides a considerably improved variational solution over the ansatz 
proposed in Ref.~\cite{garcia-march_entanglement_2016}. Nevertheless, here we stress that the ansatz becomes less accurate with increasing $N$.

\section{Energy dependence on interaction}
\label{sec:energy}

We first examine the energies of the system for different interaction-strength regimes. In the following, we start by providing an exhaustive examination of the ground states, to then study the low-energy spectrum.

\subsection{Ground-state energy for strong interactions}
\label{sec:energy;subsec:ground}

To analyze the ground-state energies, we study the energy increase of the system due to the impurity-bath interaction
\begin{equation}
    \mu(g,N) \equiv E(g,N)-E(g=0,N)\,,
\end{equation}
where $g$ simply indicates the interaction strength at which the energy is obtained
and $N$ is the number of particles employed.
In related studies, $\mu$ is usually referred to as the polaron or binding energy of the impurity~\cite{jorgensen_observation_2016}, which can be interpreted as the energy required to add the impurity to the system.
Let us emphasize that the ground state of the system is the same regardless of the nature of 
the bath, i.e. identical bosons or distinguishable particles.
We also note that we examine excited states in the next subsection.

\begin{figure}[t]
    \centering
    \includegraphics[width=0.95\textwidth]{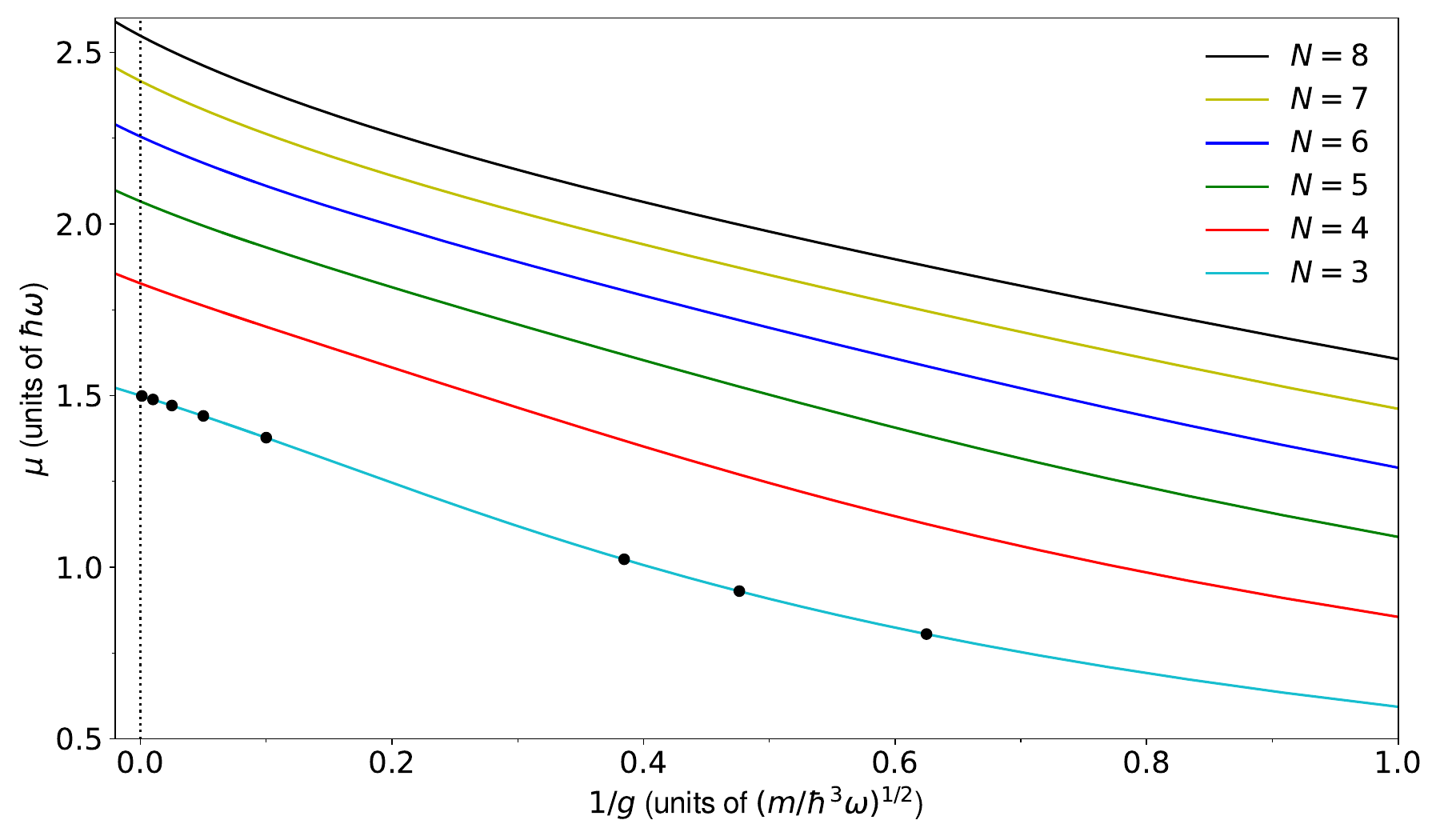}
    \caption{Ground-state binding energy $\mu$ as a function of the inverse of the interaction strength $1/g$.  We show results for three to eight particles, as indicated in the labels. The solid lines correspond to results from ED, while the black circles correspond to the results reported in~\cite{zinner_fractional_2014}. We have employed 90, 45, 30, 20, 15, and 15 HO modes for three, four, five, six, seven, and eight particles, respectively.}
    \label{sec:energy;fig:gs}
\end{figure}

We show values of $\mu$ obtained from ED calculations for up to $N=8$ ($N_b=7$) particles in figure~\ref{sec:energy;fig:gs}. We plot the results as a function of the inverse of the interaction strength to focus on the strongly-interacting regime $g^{-1}\approx 0$. We also show exact results for $N=3$ from Ref.~\cite{zinner_fractional_2014} (black circles), giving a perfect agreement with our numerical calculations.

First, and as expected, we note that $\mu$ increases with increasing $g$,  reaching non-trivial values for $g^{-1}\to 0$. Second, we find that $\mu$ also increases with $N$. We stress that by construction, $\mu(g=0)=0$ for all numbers of particles. Therefore, it requires more energy to add an impurity to a more populated and more repulsive bath, as expected.

We also report the obtained values of $\mu$ at $g^{-1}\approx 0$ in~\tref{sec:energy;tab:e_k}. We report both the energies obtained from our proposed ansatz~\eref{sec:model;eq:ansatz2} as well as the energies obtained from the simpler ansatz from Ref.~\cite{garcia-march_entanglement_2016}. We obtain a reasonable agreement between the values obtained from ED and ansatz~\eref{sec:model;eq:ansatz2}, especially for a small number of particles, showing that the proposed ansatz gives a good description of this configuration. However, and as previously mentioned, the agreement decreases with an increase in the number of particles, reaching a difference of approximately 24\% (9\% for the total energy) for $N=8$. On the other hand, we find a larger disagreement between our numerical calculations and the ansatz proposed in~\cite{garcia-march_entanglement_2016}, reaching a difference of approximately 59\% for $N=8$. This shows that the proposed ansatz \eref{sec:model;eq:ansatz2} provides a significantly improved description. Nevertheless, the discrepancies between the numerical calculations and the ansatz suggest that to describe many-body configurations we cannot rely on simple analytical examinations. We also stress that  Ref.~~\cite{zinner_fractional_2014} found that the exact value for $N=3$ in the infinite interacting limit corresponds to $\mu=3/2\hbar\omega$, in almost perfect agreement with our numerical calculations.

\begin{table}[t]\label{sec:energy;tab:e_k}
    \caption{Energies $\mu$ (in units of $\hbar\omega$) and derivatives $K$ (in units of $(\hbar^5\omega^3/m)^{1/2}$) computed for $g^{-1}\approx 0$ for different number of particles $N$. The ED results (first and fourth rows) are obtained for $g = 1000 (m/\hbar^3\omega)^{1/2}$. The second row shows the energies at $g^{-1}=0$ obtained from ansatz~\eref{sec:model;eq:ansatz2}, while the third row shows the energies obtained with the ansatz proposed in Ref.~\cite{garcia-march_entanglement_2016}.}
    \begin{indented}
    \lineup
    \item[]\begin{tabular}{@{}*{7}{l}} 
        \br
        $N$ &  3 & 4 & 5 & 6 & 7 & 8\\ 
        \mr
        $\mu_{\mathrm{ED}}$ & 1.499 & 1.826 & 2.064 & 2.253 & 2.415 & 2.549 \\ 
        $\mu_\mathrm{ansatz}$ (\ref{sec:model;eq:ansatz2}) & 1.537 & 1.948 & 2.298 & 2.608 & 2.891 & 3.152 \\
        $\mu_\mathrm{prev.}$ (Ref.~\cite{garcia-march_entanglement_2016}) & 1.667 & 2.143 & 2.596 & 3.068 & 3.552 & 4.043\\
        \mr
        $K$ & 1.145 & 1.379 & 1.538 & 1.678 & 1.834 & 1.933\\
        \br
    \end{tabular}
    \end{indented}
\end{table}

To further characterize the limit of infinite repulsion, we also report in~\tref{sec:energy;tab:e_k} the values obtained with ED for the derivative of $\mu$ at $g=0$,  
\begin{equation}
    K=-\left[ \frac{d\mu}{dg^{-1}} \right]_{g^{-1}=0}\,.
\end{equation}
This magnitude is connected to Tan's contact~\cite{barth_tan_2011} and can be used to compute
the interaction energy as $E_\mathrm{int.}=g^{-1} K$, as dictated by the Hellman-Feynman theorem. 
We obtain that $K$ increases with $N$, indicating that for large interactions (but not infinite) the interaction energy increases with the number of particles.
We also note that  Ref.~\cite{zinner_fractional_2014} reported the exact value $K=\frac{9}{\sqrt{2 \pi^3}}\approx 1.143$ for $N=3$, also in almost perfect agreement with our numerical results.

Within the range of the number of particles examined, even though $\mu$ increases with $N$, we find that this increase becomes smaller with larger baths. Indeed, we find that 
\begin{equation}
\label{sec:energy;eq:inf_e_relations}
    \mu(\infty,N+1)-\mu(\infty,N) > \mu(\infty,N+2)-\mu(\infty,N+1)\,,
\end{equation}
which can be appreciated in both \fref{sec:energy;fig:gs} and \tref{sec:energy;tab:e_k}. To provide energy estimates for systems with many particles, we fit the obtained numerical energies at $g^{-1}\approx 0$ to the function
\begin{equation}
\label{sec:energy;eq:inf_e_fit}
    \mu(\infty,N)=\Delta E \left( 1-\frac{1}{N^{b}} \right)\,,
\end{equation}
where $\Delta E$ and $b$ are parameters to determine. The best fit for our numerical results gives $\Delta E=6.7\pm0.2$ (in units of $\hbar\omega$) and $b=0.228\pm0.009$, showing that our results are well adjusted by function (\ref{sec:energy;eq:inf_e_fit}). 

The parameter $\Delta E$ gives an estimate for the value of $\mu$ for an infinite number of particles at 
the infinite repulsion limit. Because $\Delta E$ is finite, we find that 
$\mu(\infty,N)$ saturates for large $N$. Naturally, the total ground-state energy of 
the system $E$ diverges for infinite $N$.
Nevertheless, because our calculations consider only a few particles, this 
extrapolation should be taken with care and should be contrasted with a robust 
many-body calculation in the future. However, this saturation can be expected from a qualitative examination. Indeed, in baths with a large number of particles, the impuritiy's wavefunction cannot have overlap with the wavefunction of the bath. Therefore, in this scenario, adding one more 
particle to the non-interacting bath will not add any extra interacting energy to the 
impurity, resulting in the saturation of the binding energy $\mu$ for large $N$.

\subsection{Low energy spectrum}

\begin{figure}[t]
\centering
\includegraphics[width=0.95\textwidth]{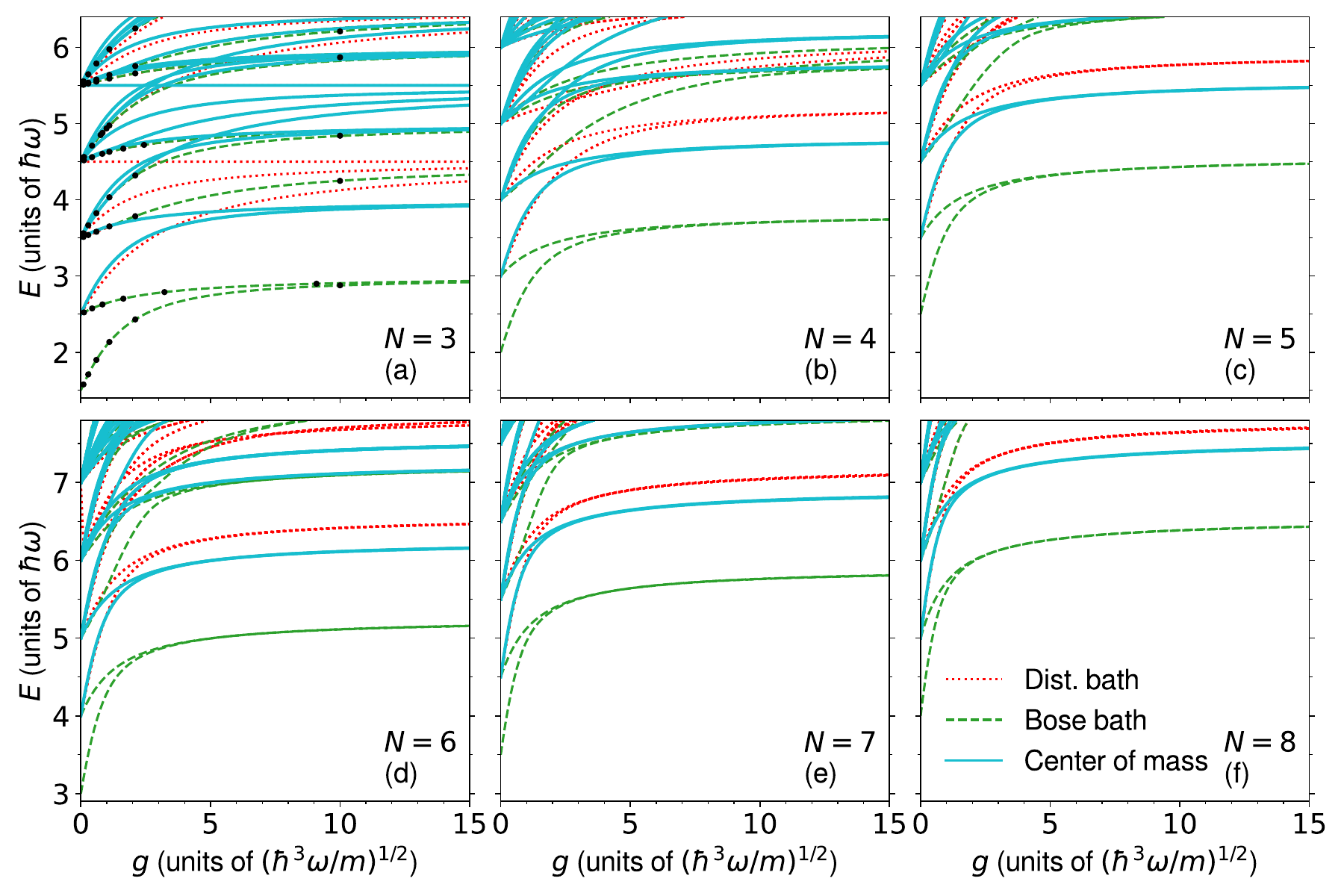}
\caption{Low-energy spectrum $E$ of three (a), four (b), five (c), six (d), seven (e), and eight (f) particles, including the 
interacting impurity, as a function of the interaction strength $g$. The lines correspond to results from ED while the black circles correspond to 
the results from~\cite{zinner_fractional_2014}. We use 90, 45, 30, 20, 15, and 15 HO modes for the three, four, five, six, seven, and eight particles, respectively. The dashed green lines correspond to the ground state and the intrinsic excitations of a system with either a bosonic or a distinguishable bath, while the dotted red ones correspond to the states that are present only with a distinguishable bath.} Finally, the solid cyan lines correspond to the center-of-mass excitations.
\label{sec:energy;fig:spectrum}
\end{figure}

Having examined the ground state, we now turn our attention to the full energy in the lower part of the spectrum. In \fref{sec:energy;fig:spectrum} we show the low-energy spectrum obtained from ED as a function of $g$ for three to eighth particles. We also show exact results for $N=3$ from Ref.~\cite{zinner_fractional_2014} (black circles), giving a perfect agreement with our numerical results. As mentioned before, the ground state of a system with a bath of non-interacting identical bosons is identical to that with a bath of distinguishable particles. However, this situation slightly changes in the excitation spectrum, where the distinguishable bath shows additional intrinsic excitations. In the figure, we show the ground state and intrinsic excitations shown by both baths with dashed green lines, while we show intrinsic excitations particular to the distinguishable bath with dotted red lines. Finally, center-of-mass (c.m) excitations are highlighted with solid cyan lines. These c.m. excitations arise from the separability of the c.m. Hamiltonian~\cite{busch_two_1998} and have an energy gap of $n\hbar\omega$ ($n=1,2,...$) with respect to the ground state or intrinsic excitations.

Naturally, in the non-interacting limit $g=0$, the energy spectrum is simply given by the HO solution $E_{n}(g=0)=\hbar\omega\sum^N_{a}\left(n_a+1/2\right)$. In contrast, for strong repulsion, the spectra saturate to non-trivial values that do not correspond to eigenvalues of the HO Hamiltonian.

In the non-interacting limit $g=0$, the energies show the expected degeneracies for $N$ distinguishable particles in a harmonic oscillator. Some of these degeneracies are lifted for $g\neq 0$, as seen in panels (a-f) of the figure. However, in the limit $g\to +\infty$ new degeneracies are introduced.

Indeed, in the strongly interacting limit, $g\to +\infty$, the ground state becomes double degenerate in all cases, with one even and one odd parity states (see right part of panels (a-f) in \fref{sec:energy;fig:spectrum}). 
In this limit, the first excitation (lowest blue lines in all panels) is a center-of-mass
excitation of the ground-state doublet and has an energy gap of $\hbar\omega$ for all values of $g$. 
Slightly above the lowest center-of-mass excitation (for $g\to\infty$), the spectrum shows an intrinsic excitation which is only present with a distinguishable bath. 
For $N=3$, this intrinsic excitation is created by a manifold of four states and reaches the 
Tonks-Girardeau energy for three particles, $E=9/2\hbar\omega$. 
For $N>3$, this intrinsic excitation is instead created by a doublet structure of opposite parity. 
Each line of the doublet is itself degenerate $D=N-2$ times.

\begin{figure}
    \centering
\includegraphics[width=0.95\textwidth]{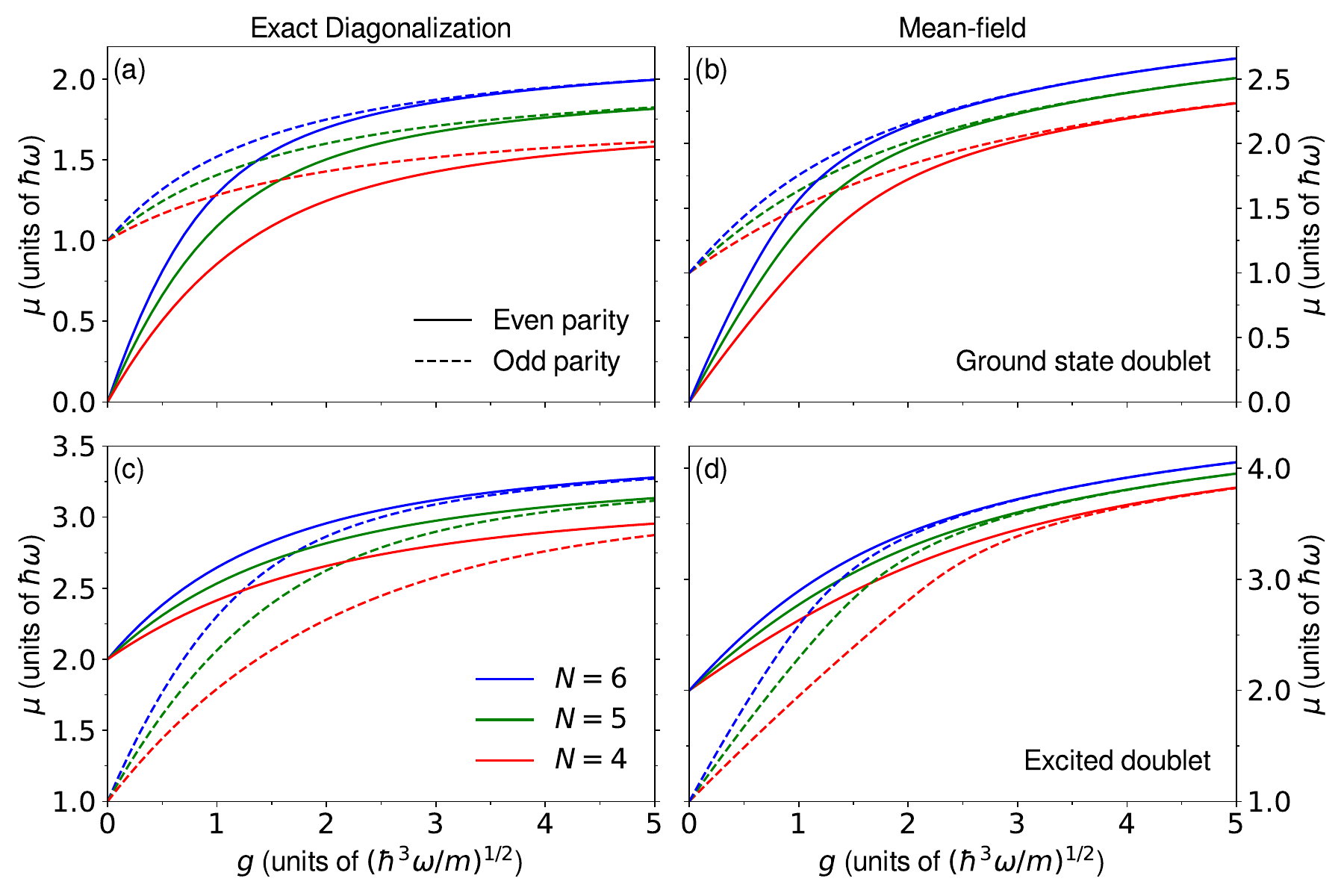}
    \caption{Low-energy $\mu$ spectrum structure for the two lowest doublets for $N=4$, $5$, and $6$. Panels (a) and (c) correspond to results obtained with exact diagonalization, whereas panels (b) and (d) correspond to mean-field solutions. Panels (a) and (b) show the ground-state doublet, while panels (c) and (d) show the first excited doublet. Solid lines correspond to states with even parity and the dashed lines to states with odd parity.}
    \label{sec:energy;fig:s.c.energy}
\end{figure}

Note that the first intrinsic excited doublet (for $N>3$) and the ground state have a similar structure. 
Therefore, to further examine these states, 
in \fref{sec:energy;fig:s.c.energy} we show the impurity energy $\mu$ for the ground state doublet (top panels) and the first
intrinsic excited doublet (bottom panels) computed by both ED (left panels) and the mean-field solution (right panels) described in \sref{sec:model;subsec:mf}. Here we note that the MF solutions provide a similar qualitative structure to that obtained with ED. We also stress again that the shown intrinsic excitations only appear for a distinguishable bath. Moreover,
for the excitations $\mu$ is defined as the difference between the energy of the excited state and the energy of
the non-interacting ground state.

As discussed in \sref{sec:energy;subsec:ground},
the ground-state energy (top panels of \fref{sec:energy;fig:s.c.energy}) in the infinite interacting limit increases with $N$. Also, the gap between the two states closes for a 
weaker interaction as the number of particles increases. Independently of $N$, each state 
of these doublets has no degeneracy, with one having even parity while the other having odd parity.

The first intrinsic excited doublet (bottom panels of \fref{sec:energy;fig:s.c.energy}) has the same behavior as the ground state: as the number of particles increases
the value of the energy in the infinite interacting limit increases and the gap between the 
two states close for weaker interactions. Each curve of the excited doublet is $N-2$ 
times degenerate, and all of the states in one curve have either even parity or odd parity. Nevertheless, we note that the excited doublets close 
their gap for a larger value of the interaction compared with the ground-state doublets. The energy difference between the two doublets in the infinite interacting limit
is larger than $\hbar\omega$. This indicates that the excited doublet gains more
energy from zero to infinite interaction than the ground state.

We can easily interpret both doublet states with a simple model in which the impurity is immersed in a double well, where the central barrier is created by the repulsive bath.
The height of the central barrier naturally depends on the number of bath particles $N_b$ and on the interaction strength $g$. The Hamiltonian of this one-particle effective model is
\begin{equation}
    \hat{H}=-\frac{\hbar^2}{2m}\frac{d^2}{d x^2}+\frac{m\omega^2}{2}x^2 +gN_b |\varphi(x)|^2\,,
    \label{sec:energy;eq:doublewell}
\end{equation}
where we can take $\varphi(x)$ as the harmonic oscillator ground-state wavefunction. Note that our mean-field approach is a
direct variation of this model, where we introduce the possibility to adapt the central barrier created by the 
bath particles. Indeed, solving the effective double-well Hamiltonian (\ref{sec:energy;eq:doublewell}) is equivalent to directly solving the MF equation (\ref{sec:model;eq:s.c.imp}) assuming that the bath's bosons are in the HO ground-state.

In this double-well model, the two lowest 
states become degenerate when the barrier becomes infinite and they have opposite parity.
For this reason, we can understand the doublet states as the ground state and the 
first excitation of the impurity in the double well. This interpretation is also useful to explain the 
excited doublet. In that case, we can understand it as an excitation of the central barrier. 
This produces a wider central barrier, whose effects are to increase the 
energy in the infinite interacting limit and to close the gap for larger values of the 
interaction.
We note that the excitation of the central barrier only makes sense for the case with distinguishable baths, as those states do not appear in the spectrum of bosonic baths.

As mentioned, the MF solutions (right panels of \fref{sec:energy;fig:s.c.energy}) can be seen as a correction of the double-well model.
Indeed, the excitations of the particle in the double well
means excitations of the impurity in the MF approach, while the excitation of
the central barrier in the double-well model corresponds to an excitation of 
one bath particle in the MF.

As previously discussed, the MF solutions provide similar results to those obtained with ED. However, one important difference is that the MF solutions do not converge to finite energy in the limit
of infinite interaction. In the same panels, we can also note that the state with
an impurity excitation also diverges for large interactions. Nevertheless, both
states degenerate for large interaction strength. 

In addition, the MF solutions close the gaps at weaker interaction strengths than the ED results and reach larger energies at the same interactions.

\section{Ground-state density profiles}\label{sec:density}
\begin{figure}[t]
    \centering
    \includegraphics[width=0.95\textwidth]{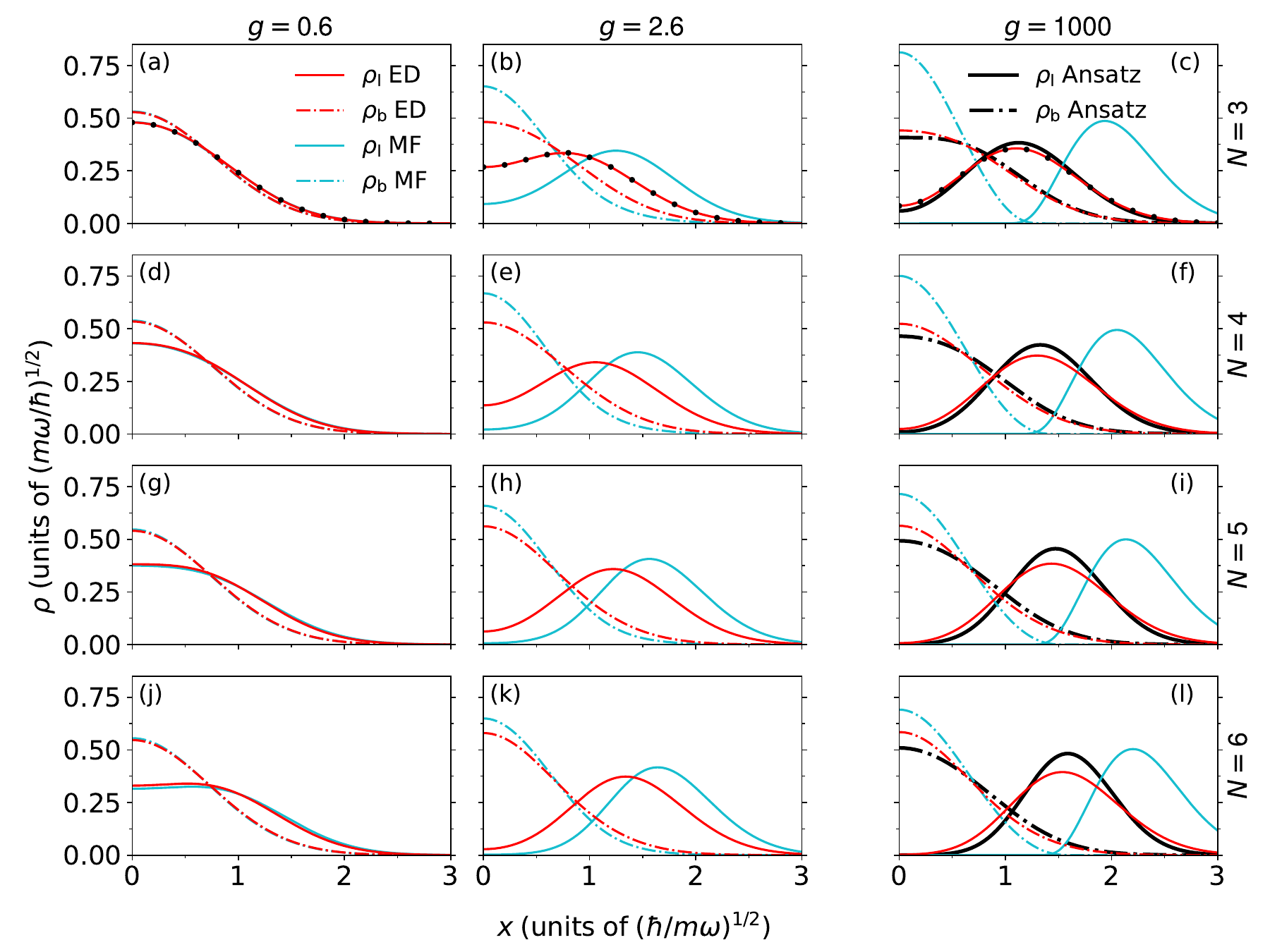}
    \caption{Density of the ground state for three (a, b, c), four (d, e, f), five (g, h, i), and six (j, k, l) particles for a selection of interaction strengths $g$. Each column corresponds to results for a chosen interaction strength $g$, as indicated at the top of the figure, with $g$ in units of $(\hbar^3\omega/m)^{1/2}$.
    The solid lines correspond to the density of the impurity ($\rho_I$), whereas the 
    dash-dotted lines correspond to the density of one of the bath particles ($\rho_b$).
    The red lines correspond to ED calculations, the cyan lines correspond to MF solutions, and the black lines in the right panels correspond to the densities predicted by ansatz~\eref{sec:model;eq:ansatz2} for infinite repulsion.
    Additionally, the black circles in the top panels correspond to the impurity's density reported in~\cite{zinner_fractional_2014} for $N=3$.
    Note that for small interactions ($g/(\hbar^3\omega/m)^{1/2}=0.6$) the MF results are almost indistinguishable from the ED results.}
    \label{sec:density;fig:dens}
\end{figure}

In this section, we examine the density profiles of the ground state. This enables us to further understand the behavior of the particles across different interaction strengths, and better explain some of the results that we found previously.

We first show the ground-state density profiles obtained from ED for three, four, five, and six particles 
in \fref{sec:density;fig:dens}. We show profiles for the impurity (solid lines) and the bath's particles
(dash-dotted lines). Note that in the ground state, the profiles of all the bath particles are equal. We
show results for a representative set of interaction strengths, where the four left panels correspond to a
weak repulsion, the four right panels to essentially the infinite repulsion limit, and the middle panels
correspond to an intermediate configuration. To contrast our calculations, we also show the corresponding MF
solutions (cyan lines) and the ansatz for infinite repulsion (black lines). We also compare with the
results from Ref.~\cite{zinner_fractional_2014} for $N=3$, which show a perfect agreement with our ED
calculations. Note that the profiles are symmetric with respect to $x$.

For $g/(m/\hbar^3\omega)^{1/2}=0$, all the particles are in the HO ground state and thus the profiles are simply Gaussian functions. As the interaction strength increases, the density of the
impurity starts developing a two-peak structure, with a minimum at the center of the trap ($x=0$).
In contrast, the bath particles develop a complementary density, with a maximum at the center of the trap.
Therefore, the bath remains at the center of the trap, acting as a barrier that pushes the impurity to the edges of the trap. This results in an effective double well-like potential for the impurity, as discussed previously.

As the number of particles increases, we observe that, for a fixed $g$, the impurity is further repelled from the center of the trap, as can be expected from a system with a larger repulsive bath.
In the same way, the perturbation of the impurity in the bath becomes smaller. 
From a MF point of view, the strength of the central barrier created by the bath is proportional to the product of the interaction 
strength and the number of bath particles. 

For weak interactions (left panels) the MF solutions give a good description, as can be expected, with profiles that are almost indistinguishable from the ED results. However, as $g$ increases, particularly when the impurity develops its two peaks structure, the MF solutions start deviating from the ED results (middle panels), to then give a poor description for very strong repulsion (right panels). Note that while
the overlap between the bath and impurity profiles for strong repulsion (right panels) is small but finite with ED, the MF profiles show an almost vanishing overlap. Nevertheless, the MF approach still is able to give a qualitative description of the strongly-interacting regime, as it correctly predicts the separation of the bath and impurity.

For strong repulsion (right panels), the ansatz~\eref{sec:model;eq:ansatz2} shows a good agreement with the ED profiles, showing that our ansatz correctly captures the main features of the wavefunctions. The ansatz shows a finite overlap between the profiles, unlike the MF solutions, with profiles that extend to larger distances.

\begin{figure}
    \centering
    \includegraphics[width=0.95\textwidth]{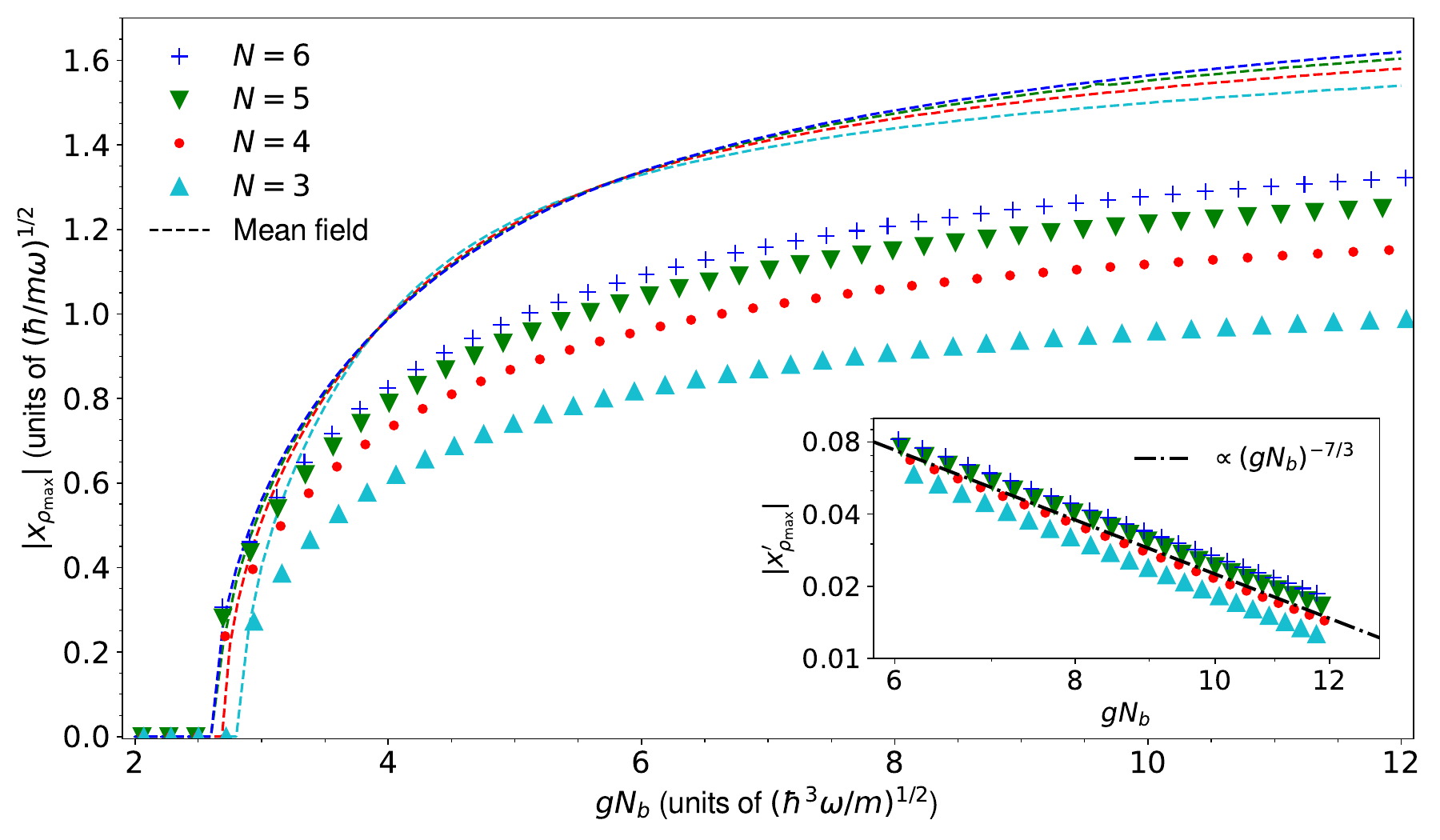}
    \caption{Position of the impuritiy's density's maximum $x_{\rho_\mathrm{max}}$ as a function of $g N_b$. The markers correspond to numerical results obtained with ED, whereas the dashed lines 
    correspond to MF solutions. The inset shows the derivative of $x_{\rho_\mathrm{max}}$ with respect to $gN_b$
    in a double logarithmic scale. The dash-dotted line in the inset is an eye guide that shows the power law of the derivatives. 
    }
    \label{sec:density;fig:maxdens}
\end{figure}

In order to further characterize the double well interpretation, in \fref{sec:density;fig:maxdens} we show the position of the maximum of the impurity's density $|x_{\rho_\mathrm{max}}|$ as a function of the product between the interaction strength $g$ and the number of bath's particles $N_b$. We show results obtained numerically with ED (markers) and the corresponding MF solutions (dashed lines) as obtained from~\eref{sec:model;eq:s.c.imp} and \eref{sec:model;eq:s.c.bath}.  We stress that, as explained before, $gN_b$ dictates the height of the effective central barrier. We also note that we consider the absolute value of $x_{\rho_\mathrm{max}}$ because the system is symmetric with respect to $x$.

While the MF solutions deviate considerably from the ED results as $g$ increases, they both show the same qualitative behavior. Indeed, both the ED and MF results predict that, for small $g$, the impurity's density 
has only one maximum at the center of the trap, and therefore $x_{\rho_\mathrm{max}}=0$ (left region in the figure). However, the impurity's density develops two symmetric 
peaks at interaction strengths larger than a critical strength $g^*$, as illustrated by the appearance of a finite $x_{\rho_\mathrm{max}}$ (right region). These two peaks develop suddenly at $g^*$, 
and thus $x_{\rho_\mathrm{max}}$ shows a discontinuous derivative. 

Interestingly, the critical point $g^* N_b$ depends weakly on the number of particles $N$, and this dependence decreases with increasing $N$. Indeed, we find that for larger baths the ED critical point converges to $g^*N_b/(\hbar^3\omega/m)^{1/2}\approx 2.5$. Therefore, the impurity shows a seemingly universal transition between being localized at the center of the trap and being expelled to its border. 
Naturally, this transition should be examined further in the future with robust many-body approaches.

This transition can be understood as a mean-field phase separation between two species~\cite{ho_binary_1996,pu_properties_1998}
, explaining why the transition point is so well captured by MF.  Using energy arguments, we note that the system has an interaction energy and potential energy due to the position of the impurity. For weak repulsion, the ground state has the
minimum potential energy and thus only the interaction energy can increase. However, for strong interactions starting from a transition point, it is energetically
favorable to have the impurity away from the center of the trap, increasing the potential energy, but 
decreasing the interaction one.

Even though the transition is well described by MF, and as explained before, the MF approach is unable to give a good description for strong repulsion (right region in the figure), showing a noticeable disagreement with the ED results. Therefore, to describe such a regime we need to rely on more robust approaches, such as ED. Here we stress that precisely the regime of strong repulsion is of interest in the study of one-dimensional Bose polarons~\cite{will_polaron_2021,will_dynamics_2023}
.

Finally, we also find that $|x_{\rho_\mathrm{max}}|$ saturates to a finite value in the limit of large repulsion $g N_b\to\infty$ with both ED and MF. To show this convergence, we show the derivative of $|x_{\rho_\mathrm{max}}|$ obtained from ED in the inset of \fref{sec:density;fig:maxdens} (note the logarithmic scale). It is easy to see that the derivative behaves as a power law. Indeed, we find that it behaves approximately as $\sim 1/(gN_b)^{7/3}$ (see dashed line). The results obtained with MF 
also saturate, but the derivate behaves as $\sim 1/(gN_b)^{10/3}$. Moreover, and as with $g^*N_b$, the saturation position depends on $N$, but this dependence decreases with increasing $N$.

\section{Conclusions}
\label{sec:concl}

In this article, we have examined the energy spectrum and density profiles of the system of an impurity immersed in a bath of a few non-interacting particles, being identical bosons or distinguishable particles. We have performed exhaustive numerical diagonalizations to examine a wide range of repulsive impurity-bath interaction strengths and baths with a set of different numbers of particles $N_b$. The system shows a rich non-trivial behavior at finite interaction strengths.
In particular, we have found that the ground-state impurity energy in the infinite repulsive
limit saturates for large $N$, showing that the energy required to add an impurity to the bath becomes independent of the size of the bath. 

We have studied both weak and strong interaction regimes. We have used a mean-field approximation in order to 
describe the weakly-interacting regime and we have proposed an ansatz for the infinite interacting limit. 
Both techniques are in agreement with our exact diagonalization calculations in their respective regimes of application.

Via the energy spectrum, we have identified a double-well behavior in the ground state doublet that closes the energy gap 
as the interaction becomes infinite. In addition, we have examined some excited states with similar behavior. 
These states can be understood with a single-particle effective model, where the impurity is 
immersed in a double well potential in which the central barrier is created by the bath particles.
We use a mean-field approach based on this simple model where the central barrier changes due to the interaction. As a consequence, it explains that the impurity is pushed away for large repulsive interactions and becomes localized at the edges of the trap.

By examining the density profiles, we find that indeed the impurity shows a transition between being localized at the center of the trap for weak repulsion, to being expelled to the border of the trap for strong repulsion. The latter corresponds to the double-well regime. Furthermore, we find that the transition manifests at a seemingly universal critical point $g^* N_b$.

Having studied the problem of a single impurity immersed in a non-interacting bath, we devise several future-related extensions of this work. One natural extension is to consider interacting baths to further study dressed impurities. On top of adding interactions to the model studied here, of particular interest would be to study baths composed of two-component interacting fermions, as recently studied in homogeneous configurations \cite{pierce_few_2019,rammelmuller_magnetic_2023}. Consideration of two impurities is also of interest \cite{theel_crossover_2023,petkovic_mediated_2022}
, especially due to its connection to the bipolaron problem \cite{camacho-guardian_bipolarons_2018,will_polaron_2021}. In this direction, ED calculations would enable us to easily examine correlations between impurities.
In addition, our developed ED techniques can be extended to study dynamics, which could be employed to study the polaron formation~\cite{will_dynamics_2023} 
and quench dynamics~\cite{mistakidis_effective_2019,mistakidis_many-body_2020}. Finally, the findings of this work could be further examined in the future for larger baths using robust many-body approaches, such as with quantum Monte-Carlo.

\section*{Acknowledgements}
We thank J. Martorell for the useful discussions and for carefully reading our manuscript. We thank G. Astrakharchik for 
useful comments and discussions. We also thank A. Volosniev 
for providing us the data from~\cite{zinner_fractional_2014}.

\paragraph{Funding information}

This work has been
funded by Grant No. PID2020-114626GB-I00 from
the MICIN/AEI/10.13039/501100011033, and by Grant FPU20/06174 from the Ministry of Universities of Spain. F.I. acknowledges funding from EPSRC (UK) through Grant No.
EP/V048449/1 and from ANID (Chile) through FONDECYT Postdoctorado No. 3230023.

\appendix
\section{Analytical ansatz calculations}\label{app:analytical}

In the following we examine ansatz~\eref{sec:model;eq:ansatz2} in more detail. First, we
write explicitly how we determine the weight of each contribution.
Then, we show the ansatz for four and five particles, and finally, we 
present a table with the values of $\sigma_I$ and $\sigma_b$ obtained up to $N=8$.\\ \\
In order to obtain the weights of each component of the ansatz, 
we start from the simplest version, i.e., the superposition of~\eref{sec:model;eq:ansatz_N3_1} and ~\eref{sec:model;eq:ansatz_N3_2}. This reads 

\begin{equation}\label{sec:app;eq:ansatz_N3}
    \Psi(x_I,x_i)=\alpha \left[(x_I-x_A)(x_I-x_A)-c|x_I-x_A||x_I-x_A|  \right]e^{-(x_I^2+x_A^2+x_B^2)/2}\,,
\end{equation}
where we use $c$ as a relative weight between each
two components. The energy is obtained by integrating 
$\Psi(x_I,x_i)\hat{H}\Psi(x_I,x_i)$ and normalizing the constant $\alpha$. Therefore, for~\eref{sec:app;eq:ansatz_N3} we obtain
\begin{equation}
    E=\frac{57 \pi  c^2+102 \sqrt{3} c+38 \pi  c+57 \pi }{6 \left(\pi  (c (3 c+2)+3)+6 \sqrt{3} c\right)}\,.
\end{equation}
By minimizing for $c$, we obtain 
$E\simeq 3.06916$ with $c=1$. The solution with $c=1$ is also obtained 
numerically when we introduce the $\sigma$ parameters.

For a larger number of particles we also obtain that all the wavefunction components have the same weight. In~\tref{sec:appendix;tab:sigma} we show the variational constants $\sigma$ for values of $N=3$ to $8$ that minimize the total energy corresponding to ansatz~\ref{sec:model;eq:ansatz2}. We find
that $\sigma_b$ is roughly independent of $N$, whereas $\sigma_I$ shows a stronger dependence on $N$

To further illustrate the ansatz, we present the explicit expression for four and five particles. The expression for three particles is in the main text in \eref{sec:energy;eq:ansatz_3p}.
For four particles the ansatz reads
\begin{eqnarray}
\fl |\Psi_4,\sigma\rangle=
\exp\left[ -x_I^2/2\sigma_I^2-\left(x_A^2+x_B^2+x_C^2\right)/2\sigma_b^2 \right]
\left[( x_I-x_A)  | x_I-x_B|  | x_I-x_C|  \right. \nonumber\\
\fl +(x_I-x_B) |x_I-x_A| | x_I-x_C|   + (x_I-x_C) |x_I-x_A| | x_I-x_B|   \nonumber \\
\fl \left.+(x_I-x_A) (x_I-x_B) (x_I-x_C) \right]\,,
\end{eqnarray}
while for five particles it reads
\begin{eqnarray}
\fl |\Psi_5,\sigma\rangle=
\exp\left[-x_I^2/2\sigma_I^2-\left(x_A^2-x_B^2-x_C^2-x_D^2\right)/2\sigma_b^2\right]\left[| x_I-x_A|  | x_I-x_B|  | x_I-x_C|  | x_I-x_D|\right. \nonumber\\
\fl +(x_I-x_A) (x_I-x_B) | x_I-x_C|  | x_I-x_D| + (x_I-x_A) (x_I-x_C) | x_I-x_B|  | x_I-x_D| \nonumber \\
\fl +(x_I-x_B) (x_I-x_C) | x_I-x_A|  | x_I-x_D|  +(x_I-x_A) (x_I-x_D) | x_I-x_B|  | x_I-x_C| \nonumber \\
\fl +(x_I-x_B) (x_I-x_D) | x_I-x_A|  | x_I-x_C| +(x_I-x_C) (x_I-x_D) | x_I-x_A|  | x_I-x_B| \nonumber \\
\fl \left.+(x_I-x_A) (x_I-x_B) (x_I-x_C) (x_I-x_D)\right]\,.
\end{eqnarray}
\begin{table}[t]\label{sec:appendix;tab:sigma}
    \caption{Values of the parameters $\sigma$ that minimize the energy of ansatz~\eref{sec:model;eq:ansatz2}. These values are obtained numerically.}
    \begin{indented}
    \lineup
    \item[]\begin{tabular}{@{}*{7}{l}} 
        \br
        $N$ &  3 & 4 & 5 & 6 & 7 & 8\\ 
        \mr
        $\sigma_I$ & 0.918427& 0.853892& 0.805508 & 0.767799 & 0.737288 &  0.711888\\ 
        $\sigma_b$ & 0.946936 & 0.932079 & 0.927925 & 0.927379 & 0.928215 & 0.92959  \\
        \br
    \end{tabular}
    \end{indented}
\end{table}

\clearpage
\bibliography{biblio_impurity}

\end{document}